\documentstyle[pre,multicol,aps,epsf]{revtex}
\begin{document}
\draft

\title
{\Large \bf 
Coherence resonance  near  blowout bifurcation in nonlinear dynamical systems 
}

\author{ Bambi Hu$^{1,2}$, Changsong Zhou$^1$ }

\address 
{ $^1$ Department of Physics and Center for Nonlinear Studies, 
Hong Kong Baptist University, Hong Kong, China\\ 
$^2$ Department of Physics, University of Houston, Houston, Texas 77204}

%\date{}
\maketitle

\begin{abstract}

Previous studies have shown that noise can induce coherence resonance
in some nonlinear dynamical systems close to a bifurcation  of a  
periodic motion, such as in excitable systems.  
We demonstrate that coherence resonance can be observed in systems 
close to a {\sl blowout bifurcation}. 
It is shown that for dynamical systems with an invariant subspace in which there
is a phase-coherent chaotic attractor, the interplay among the oscillation of local transverse
stability, noise and nonlinearity can lead to coherence resonance phenomenon.
The mechanism of coherence resonance in this type of system is different  from 
that in previously studied systems.

\end{abstract}
\pacs{ PACS number(s): 05.40.-a, 05.45.-a}

\begin{multicols}{2}

\section{Introduction}

The behavior of nonlinear dynamical systems subjected to noise
has been an interesting subject of recent investigation. A great 
deal of work has been devoted to stochastic resonance~\cite{ghjm}, where an optimal
amount of additive noise can generate the maximal response of the 
system to a weak  external periodic or aperiodic signal.  

Coherent motion can be induced {\sl purely by noise} in  some  dynamical 
systems {\sl without an external signal}.   
In nonlinear dynamical systems near the onset of bifurcations of periodic orbits, 
the periodicity is visible even before the bifurcation actually occurs if there 
is noise present, a phenomenon called {\sl noisy precursor} of the bifurcation~\cite{w}.
Similar  phenomenon has also been observed  in excitable systems~\cite{hdnh}, 
such as in various neural models~\cite{pk,l,lnk} and laser system~\cite{dkl}, 
in the fixed point regime close to a saddle-node bifurcation of a periodic orbit. 
 A  feature  common to the systems close to the bifurcation of a periodic orbit is that during the 
relaxation to  stable orbit  below the bifurcation, the transient possesses the periodicity 
above  the bifurcation, and the effect of external noise is to continually kick the system off of 
the stable orbit, so that the transient behavior displays coherent motion. 

More interestingly, recent investigations have shown that the coherence of the noise-induced
motion achieves a maximum at an optimal noise intensity~\cite{hdnh,pk,l,lnk,dkl,nss}. 
For example in an excitable system,   
when the system is kicked
away from the fixed state to  overcome a certain threshold, it will come 
back to the fixed point  only after a large excursion (noise-induced limit 
cycle, or spike in neural systems). 
When noise is weak, the system is rarely
excited, and the motion is quite irregular.  Increasing noise kicks the system
over the threshold  more often, and the system fires  more and more spikes. The 
interspike interval becomes the most regular at an optimal noise level. 
After that, too 
high level of noise  distorts  greatly the near limit cycle, rendering the
motion irregular again. 
Similar to the conventional stochastic
resonance, this phenomenon of resonance without an external signal  is called {\sl coherence
resonance} (CR).
In a recent paper,  Ohira and Sato  showed that  a simple two state  
model with time delay  can display CR~\cite{os}. Indeed, the time delay
introduces an intrinsic  periodic oscillation into the system: if noise
induces a spike at a certain moment, another spike is mostly expected 
after the time delay.  
More recent work has also shown noise-enhanced synchronization~\cite{hyps,hsg,nscm}
and array-enhanced CR~\cite{hz} in coupled or extended excitable systems.

It is interesting and practically meaningful to see whether CR exists 
in other type of system.
In this  paper, we demonstrate  CR in noisy  {\sl chaotic} dynamical systems close 
to a {\sl blowout bifurcation}~\cite{abs,pst} which occurs  
in dynamical systems    with an invariant 
subspace ${\bf S}$ in the phase space.  The transverse stability of the
subspace is determined by the motion within the subspace. When the largest
transverse Lyapunov exponent $\Lambda$ is negative, ${\bf S}$ is stable; 
while unstable when $\Lambda$ is positive, and the critical point of the 
transverse stability is the blowout bifurcation point. However, the local stability 
of the subspace may fluctuate greatly when the motion within the subspace 
is chaotic. The finite time Lyapunov exponent $\Lambda_T$ 
measuring  expansion or contraction of a transverse perturbation during
a period of finite time $T$  may oscillate greatly around the average value $\Lambda$. 
%For many chaotic or stochastic motions  where the correlation decreases 
%quickly enough, this fluctuation assumes an asymptotic Gaussian 
%distribution for $T\to \infty$. 
This fluctuation of local stability  
can lead to interesting and unusual behaviors, such as 
bubbling~\cite{abs}, on-off intermittency~\cite{pst}  
and additional complexity in the system  by unstable dimension variability~\cite{lyc}. 
When there is a quasiperiodic torus in ${\bf S}$, 
the system may undergo a transition to strange nonchaotic attractors via blowout bifurcation~\cite{yl}.

\section{System and results}

CR observed in this type of system is the most appreciable when the motions within
the subspace are oscillations with property of being phase-wise, although it may be chaotic
in the amplitude. A typical example of this type of motion is the R\"{o}ssler chaotic attractor [Fig.~1].  
In this work, we focus on  the case  where  there is  such a motion in the subspace,
 and study the system behavior in the presence 
of noise. Since the noise prevents the dynamics
from approaching the subspace indefinitely,          
the system may be repelled far away from the subspace due to
the local instability.  Previous investigations on the effects of noise in this type
of system  focused on the change  of the
universal behavior of the laminar phase distribution~\cite{cl} of the  
blowout motion in the context of on-off intermittency.
Here we focus on the coherence of the  blowout motion.   
Our  results will show that, with the increase of the  
noise level, the output (e.g. the distance from the subspace) displays
increasing coherence till too high level of noise dominates the dynamics 
and destroys the coherence, exhibiting  typical CR phenomenon. To
illustrate our findings, we consider the case where there is a 
R\"{o}ssler chaotic attractor in the invariant subspace,
\begin{eqnarray}
\dot{x_1}&=&\alpha(-x_2-x_3),\\
\dot{x_2}&=&\alpha(x_1+ax_2),\\
\dot{x_3}&=&\alpha(0.4+(x_1-8.5)x_3),\\
\dot{y}&=&[b(x_1-\bar{x}_1)+c]\sin y -y +\sigma \xi, 
\end{eqnarray}
where $\alpha$ controls the time scale of the R\"{o}ssler system, and
$\bar{x}_1$ is the time average of $x_1$. Clearly $y=0$ defines the invariant
subspace in the absence of noise $\xi$ which is a  Gaussian  
white noise with $\langle \xi (t) \xi (t^{\prime})\rangle=\delta (t-t^{\prime}).$ 
The introduction of $\bar{x}_1$ in Eq.~(4) is for the  convenience 
of discussion, because the  largest transverse Lyapunov exponent
\begin{equation}
\Lambda=\lim_{t\to \infty} \frac{1}{t}\int_0^{t}[b(x_1-\bar{x}_1)+c-1]d\tau 
=c-1,
\end{equation}
 and the  transverse stability is only controlled by the  parameter $c$. 
Eqs.~(1-4) are modified version of a physical model of superconducting 
quantum interference device (SQUID)~\cite{zmb}. Similar model with a torus in the subspace
has been studied in the context of nonchaotic strange attractors~\cite{yl}.  
In fact, the specific form of the  nonlinearity in Eq.~(4)   is of no importance for 
the CR phenomenon.  It serves to keep the system bounded.

We note that for parameter $a=0.15$,  the R\"{o}ssler 
system possesses a chaotic attractor with
some degree of phase coherence [see Fig.~1(a)]: the cycling  time $T_R$   
has a rather sharp distribution, or  equivalently,
there is a pronounced peak at frequency $\omega_0$ on the broadband 
spectrum of the chaotic signal.  Phase synchronization~\cite{rp1,opr}  
and lag synchronization~\cite{rp2} occur in coupled  R\"{o}ssler  
systems due  to this phase coherence.
For the parameter $c$  near  
the blowout bifurcation point $c=1.0$, the local stability 
undergoes large chaotic  oscillations. The subspace is only  temporally 
attracting in about half period of the loops  when $b(x_1-\bar{x}_1)+\Lambda<0$, 
while is temporally repelling in the other half period when 
$b(x_1-\bar{x}_1)+\Lambda>0$. 
In the following, we study system behavior in the presence of noise both below and above the 
blowout bifurcation $c=1.0$, with parameters $\alpha=0.1, a=0.15,
b=\sqrt{0.02}$.

\begin{figure}
\epsfxsize=8.5cm
\epsfbox{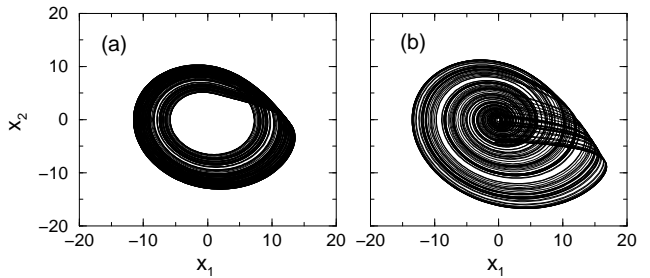}
%\centerline{\epsfig{figure=Fig1.eps,height=4cm,angle=0}}
\vspace{0.5cm}
\narrowtext
\caption{Chaotic attractors of R\"{o}ssler system with different parameters (a) a=0.15 
and (b) a=0.25.
}
\end{figure}

\subsection{Below the blowout bifurcation: $c<1$}

Typical behaviors of the output $|y|$ in the presence of noise with  different levels 
are  shown in Fig.~2 for $c=0.9$.  
 When the noise is very weak, the system stays
very closely to the invariant space $y=0$, giving no appreciable output
 [Fig.~2(a)].
If the noise is larger  than a certain level, the system begins to produce
large output, but the signal is quite 
irregular [Fig.~2(b)]. 
At a certain range of noise levels, the output become very regular; 
it is almost periodic, with the amplitude fluctuating only slightly [Fig.~2(c)].   
However, when noise  goes to even higher levels, it begins to deform the 
near periodic signal, and degrades the regularity [Fig.~2(d)].  The response 
property of the system to additive noise exhibits typical feature 
of CR: the system motion becomes the most coherent at an optimal
noise level.    

%\begin{figure}
%\epsfxsize=8.5cm
%\epsfbox{Fig2.eps}
%\vspace{0.5cm}
%\narrowtext
%\caption{Typical behavior of noise induced blowout  motion (left panel) 
%at different noise levels $\sigma$.  The  right panel is  the corresponding 
%semilog plots of the same quantity. For the purpose of clear illustration of the  
%boundary moving with the noise level, we use the 
%same scale in the right panel.  The dotted line in the upper plot 
%in the right panel indicates the noise level. (a) $\sigma=10^{-13}$, 
%(b) $\sigma=10^{-11}$, (c) $\sigma=10^{-2}$, and (d) $\sigma=10^{-0.2}$. 
%}
%\end{figure}

To characterize the degree of coherence, we compute the spectra of the
output signals $|y|$. It has a pronounced peak  at the same frequency $\omega_0$
as the driving signal $x_1$. Fig.~3(a) shows the peak height $p_m$ at $\omega_0$
as a function of the noise level $D_1=\log_{10} \sigma$. Below a certain
value of $D_1$, $p_m$ increases  exponentially  and quickly; then it comes 
to a slow increasing region covering many orders of the noise level.  $p_m$ reaches
a maximal value and decreases at higher noise levels.

In the following, let us look into the mechanism of  CR in the system.   
In the absence of noise, the subspace is stable, and from a random initial
condition, $|y|$ decreases exponentially on average as $|y|\sim \exp(\Lambda t)$, but
with a local chaotic oscillation. The additive noise, however, prevents the system from
approaching the subspace much deeper than the noise level, and the  trajectory can be repelled 
away from the noise level to produce relatively large output. Thus the noise
sets a reflection boundary to the transverse motion,  
as seen in the corresponding 
semilog plots of the output in the right panel of Fig.~2. 
Due to this boundary, 
the transverse motion is attracted to and 
repelled away from the noise level alternately when the motion within the subspace comes into 
the local stable region $b(x_1-\bar{x}_1)+\Lambda<0$ and local instable region 
$b(x_1-\bar{x}_1)+\Lambda>0$.  The  intrinsic coherence of the chaotic 
motion within the subspace is thus manifested by the noise-induced  blowout motion.  
This boundary effect of noise 
is quite different from that in excitable systems and the systems with delay, 
where noise acts to kick the system over a threshold.  

\begin{figure}
\epsfxsize=8.5cm
\epsfbox{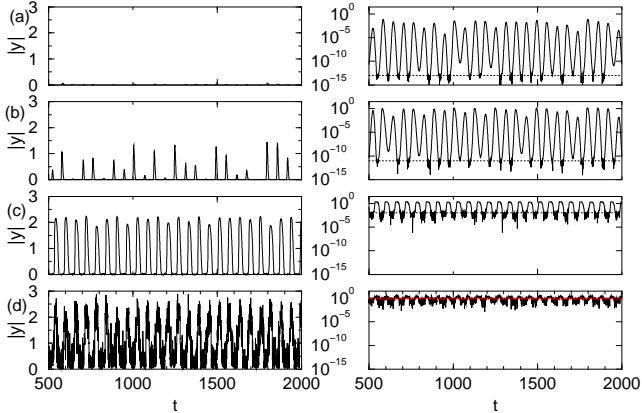}
\vspace{0.5cm}
\narrowtext
\caption{Typical behavior of noise induced blowout  motion (left panel)
at different noise levels $\sigma$.  The  right panel is  the corresponding
semilog plots of the same quantity. For the purpose of clear illustration of the
boundary moving with the noise level, we use the
same scale in the right panel.  The dotted lines 
in the right panel indicate the noise level. (a) $\sigma=10^{-13}$,
(b) $\sigma=10^{-11}$, (c) $\sigma=10^{-2}$, and (d) $\sigma=10^{-0.2}$.
}
\end{figure}

The  calculation of $p_m$  is not easy  
from solving the complete  stochastic system in Eq.~(4)  which is  driven multiplicatively 
by chaotic signal $x_1$. 
We  choose to estimate it with  more intuitive  but quite accurate approximations 
based on the above  observation that the effect of  noise can be modeled by a reflection boundary.  
For weak enough 
noise, $|y|$ keeps small, and above the noise level $|y|\gg \sigma$, the dynamics is
governed by the linear approximation $ \dot{y}=[b(x_1-\bar{x}_1)+\Lambda]y$, which  gives 
\begin{equation}
|y|=\exp\left\{\int_0^t[b(x_1-\bar{x}_1)+\Lambda]d\tau\right\}. \label{out1}
\end{equation}
Now it is not surprising that the spectra  of $|y|$ have a peak  at the same frequency $\omega_0$ as $x_1$. 
Note the intrinsic coherence of the chaotic signal $x_1$,  
we can roughly  approximate $x_1$ by a periodic signal with frequency $\omega_0$.  
Take into account the boundary of the noise, the right panel of Fig. 2 suggests that 
we may approximate $\int_0^t[b(x_1-\bar{x}_1)+\Lambda]d\tau\approx A(\cos\omega_0 t+1)+D$, 
and it follows from 
 Eq.~\ref{out1} that  
\begin{equation}
|y|\approx\exp[A(\cos\omega_0 t+1)+D] \label{apx1} 
\end{equation}
with a proper shift of the time origin.  
The amplitude $A=[b(\bar{x}_m-\bar{x}_1)+\Lambda \pi]/\omega_0$, where  $\bar{x}_m$
is the average of the maxima of $x_1$. $2A$ is the average of the  maximal  
magnitude that  $|y|$  can depart  from the  boundary at the noise level $D=\ln \sigma$ 
during a cycle of $x_1$.
With the above system parameters,  we  numerically estimate $\bar{x}_m=11.42, 
\bar{x}_1=0.1324$ and $\omega_0=1.035\alpha$, thus giving $A=12.40$.
As the noise level  increases, the boundary is moving to a higher order [Fig.~2]. 
If 
$-D\le 2A$, the maximal value of $|y|$ comes to the order of unit, and the system
begins to produce relatively large output, and the nonlinearity begins to set in. 
However, due to chaotic fluctuation of the
amplitude  of $x_1$, the maxima of 
$|y|$ may not be as large  during those small cycles of $x_1$[Fig.~2(b)],
and the behavior is quite irregular, as is familiar in the context of on-off intermittency. 
With the boundary going to even higher order, 
the system  can also produce quite large output $|y|$
during those small cycles of $x_1$ and 
the amplitude is confined  to some saturated values  by the nonlinearity [Fig.~2(c)]. 
The system now performs very coherently. In this 
nonlinear regime, Eq.~\ref{out1} based on linear approximation is no longer valid.  
However, the fact that $|y|$ always begins to rise from the noise level when $x_1$ comes into
local instable region $b(x_1-\bar{x}_1)+\Lambda>0$ shows that the blowout motion still possesses
the frequency $\omega_0$  in the strong nonlinear regime. In addition,     
Fig. 2(c) suggests that we can still approximate $\ln |y|$ by a periodic function 
whose amplitude is  
between the boundary of noise level and the confinement of the nonlinearity, i.e.,  
\begin{equation}
|y|\approx\exp[B(\cos\omega_0 t+1)+D]. \label{apx2}  
\end{equation}
Now  the amplitude $B=(D_m-D)/2$, where $D_m=\ln \max(|y|)\approx 0.8$ with the above 
system parameters.

With the  approximations in Eq.~\ref{apx1} and Eq.~\ref{apx2}, we can estimate the peak 
height at $\omega_0$  as a function of $D$ 
in different  dynamical regimes as
\begin{eqnarray}
p_m&=&\left [\frac{\omega_0}{2\pi}\int_{-\pi/\omega_0}^{\pi/\omega_0}|y|\cos\omega_0 t 
\; dt\right]^2\nonumber\\
   &\approx&\left\{ 
\begin{array}{ll}
F^2(A)e^{2A}e^{2D}, &  -D\ge 2A,\\
F^2(B)e^{2B}e^{2D}, &  -D< 2A.
\end{array}
\right. \label{analy} 
\end{eqnarray}
Here, $F(A)=\frac{1}{2\pi}\int_{-\pi}^{\pi}\cos t \; \exp(A\cos t)\;dt$. The analytical result
explains the exponential increase of the peak height in the linear regime observed 
in numerical simulation 
[Fig.~3(a), cycles].
With $A=12.80$, Eq.~\ref{analy} fits the exponential region very well 
[Fig.~3(a), solid line]. Note that the fitting  parameter $A=12.80$ is quite close 
to  $A=12.40$ estimated from the system parameters, showing that the above  simple approximations  
give  a good account for the system behavior.  The 
analysis also reproduces qualitatively  the result in the nonlinear regime. 
The discrepancy in the crossover
region ($-D\sim 2A$) is due to the fact that in this region the maximal value of $|y|$ is not saturated and 
$D_m=0.8$  used in fitting overestimates 
the maximal values of $|y|$ in this region.  
One should note that in the above analysis, the specific form of nonlinearity is of no importance, 
and only the confinement property of the nonlinearity is employed, indicating that the phenomenon 
is universal in this type of systems.   
Now we understand that  CR occurs in the system due to the interplay among  the
chaotic while somewhat coherent oscillation of the local stability of the subspace, the confinement 
of the nonlinearity to the transverse motion and the boundary effects of the 
additive noise. 
This  mechanism is quite different from that in   previously investigated 
systems~\cite{hdnh,pk,l,lnk,dkl,nss,hyps,hsg,nscm,hz,os}.

\begin{figure}
\epsfxsize=8.5cm
\epsfbox{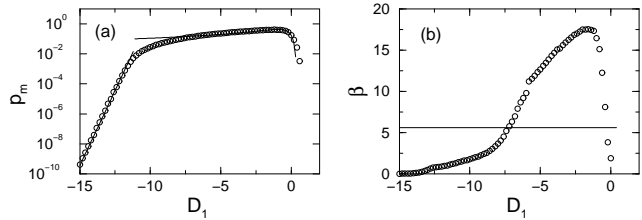}
\vspace{0.5cm}
\narrowtext
\caption{ Illustration of coherence resonance below the blowout bifurcation. 
(a) Peak height $p_m$ as a function of noise
level $D_1=\log_{10} \sigma$. The solid line is the analytical estimation with $A=12.80$.
(b) Coherence factor $\beta$ as a function of noise level. The solid line is the result 
of the chaotic signal $x_1$.   
}
\end{figure}

Based on the above understanding of the behavior, we can characterize CR by another quantity,
the ``coherence factor'' defined by the relative fluctuation of the output amplitude, e.g. 
the time average of the amplitude $A_y$ divided by its standard deviation
\begin{equation}
\beta=\langle A_y\rangle/\sqrt{{\hbox{Var}}(A_y)}.
\end{equation} 
For relatively weak noise, $|y|\gg \sigma$ most of the time,  and $|y|$ is smooth and 
$A_y$ is well defined. For relatively
large noise, noise-induced short time fluctuation of $|y|$ can be rather  strong. In numerical simulations
$A_y$ is defined as follows: firstly  a smooth  series $|y|_m$ is obtained from 
the noisy function $|y|$  using a  moving average  method, then   
$A_y$ is taken as  the value of $|y|$ at the moment when $|y|_m$ is maximal.   
This definition captures the noise-induced short time fluctuations of $|y|$ at large noise levels.   
The result is shown in Fig.~3(b) (cycles). $\beta$ reaches  a maximum and decreases 
quickly when 
too high level of noise begins to dominate the fluctuation of the amplitude.  
In linear and weak nonlinear region ($D_1<-7.5$), the chaotic fluctuation of the amplitude of $x_1$
is augmented by the exponential relationship between $|y|$ and $x_1$, as seen in Eq~(\ref{out1}), so that 
the coherence degree  of $|y|$ is lower than that of $x_1$,  as can be seen by 
the comparison of the coherence   factors of $|y|$ (cycles) and $x_1$
(solid line).  
 It is very
interesting to see that in the nonlinearity dominant region, 
the coherence of the noise-induced motion is much  higher than the intrinsic coherence of $x_1$ in
a wide range of the noise level, because the confinement of the nonlinearity smoothes 
the fluctuation of the amplitude of $|y|$. 
Thus the combination of the noise and nonlinearity enhances greatly 
the intrinsic coherence.  Such  pronounced CR phenomenon is able to be observed when
other chaotic attractors possessing similar intrinsic coherence 
are within the subspace, such as the electronic circuit in Ref.~\cite{hcp} and 
the hybrid laser system in Ref.~\cite{mf} or the ecological system in Ref.~\cite{bhs}.    
It is also clear that CR will occur for periodic and quasiperiodic motion within the subspace, i.e.,
the peaks  in the spectra possess  a maximal value at an optimal noise level.   

In the R\"{o}ssler system, the topology of the chaotic attractor changes if $a$ is large than
0.21: there are large and small
loops [see Fig.~1(b), a=0.25]. This topology is quite typical in many low dimensional  chaotic 
systems. Now, due to the large fluctuation of the amplitude and returning time,  there are no 
pronounced peaks  in the broadband spectra of the chaotic signals.  
However, CR is still observable. Fig.~4 shows the coherence factor $\beta$ for 
$|y|$ and $x_1$. Again we see the enhancement of the intrinsic coherence by noise, albeit with less
intensity. 
Similar behavior should be  observable in  general systems of this  type.

\begin{figure}
\epsfxsize=7cm
\epsfbox{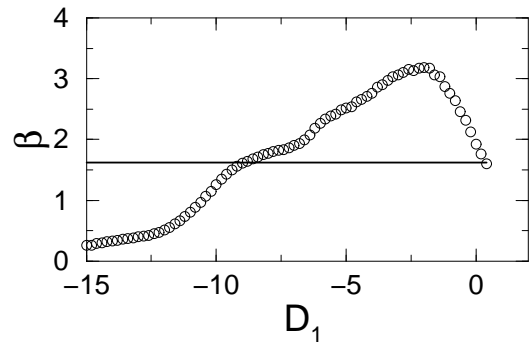}
\vspace{0.5cm}
\narrowtext
\caption{ Coherence factor $\beta$ as a function of the noise level for chaotic motion 
with funnel R\"{o}ssler attractor in Fig.~1(b).   
}
\end{figure}

\subsection{Above the blowout bifurcation: $c>1$}

Typical behavior of the system above the blowout bifurcation is similar to that below the 
bifurcation.  The difference is that, for $c>1$, the subspace is transversely unstable, and blowout motion
has already existed even without noise, and the confinement of the nonlinearity has already 
taken place. 
For $c<1$, the dynamics can always access  the noise level, 
however weak it may be; while for $c>1$, the dynamics can only come to the subspace no closer than 
$|y|\sim \exp(-2A)$, where $A=[b(\bar{x}_m-\bar{x}_1)-\Lambda \pi]/\omega_0$ for the chaotic attractor
in Fig.~1(a). As a result, weak noise with $-D>2A$ will have no discernible effects on the system 
behavior. However,  stronger noise will act as a reflection boundary similar to the case below 
the blowout bifurcation, preventing the system from approaching 
the subspace to the closest level $|y|\sim \exp(-2A)$. 
The system behavior is now very similar to that in the nonlinear regime below the bifurcation.
This analysis is demonstrated by numerical simulations with $c=1.1$  
in Fig.~5. It is seen that both $p_m$ and $\beta$ keep unchanged for very weak 
noise, and the coherence increases once the noise becomes effective. While $p_m$ increases monotonically  till 
the noise dominates the dynamics and destroys  the coherence of  the blowout motion, $\beta$ also exhibits another
peak at rather weak noise level ($\sigma\sim 10^{-10}$). This behavior is related to the topology of the 
chaotic attractor in the subspace. As seen in Fig.~1(a), the attractor always flips to the smallest loops from the largest ones.
For weak noise  without significant effects, the maxima of $|y|$ associated 
with those small loops  have relatively small 
values,  which  contributes  mainly to the fluctuation of the amplitude of $|y|$.
The broad distribution of the amplitude is clearly illustrated by the return map of the maxima of $|y|$
for $\sigma=10^{-14}$  in Fig.~6. 
For $\sigma\sim 10^{-10}$, the system has access to the  noise level when $x_1$ cycles along the largest loops.  
The reflecting  property of the noise  has the effect to  increase  greatly the  maximal values  of $|y|$  
for   those smallest  loops  following those largest ones, while only slightly for   others. 
Most of the maxima are confined to a small neighborhood around $A_y=2.0$, as seen by the crosses in Fig. ~6. 
 This 
reduces the fluctuation of the amplitude of $|y|$ and  enhances the coherence greatly. For intermediate noise 
level $\sigma\sim 10^{-7}$, the fluctuation  becomes a little larger again  
 when the maxima of  $|y|$ are  pushed to slightly larger values by 
the reflection  of the noise. Further increase of noise pushes  
more maxima of  $|y|$ to fluctuate slightly around a saturated value 
till too high level of noise destroys the coherence, resulting in  another peak.
The return map close to the peak is shown in Fig.~6 for $\sigma=10^{-2}$. 

\begin{figure}
\epsfxsize=8.5cm
\epsfbox{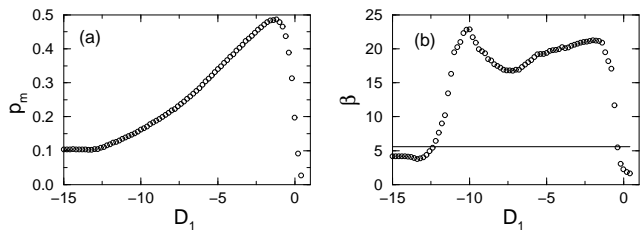}
\vspace{0.5cm}
\narrowtext
\caption{ Illustration of coherence resonance above the blowout bifurcation.
(a) Peak height $p_m$ as a function of noise
level $D_1=\log_{10} \sigma$. Unlike  Fig.~3(a), linear scale is used for $p_m$ here.
(b) Coherence factor $\beta$ as a function of noise level. The solid line is the result
of the chaotic signal $x_1$.
}
\end{figure}

\begin{figure}
\epsfxsize=8.5cm
\epsfbox{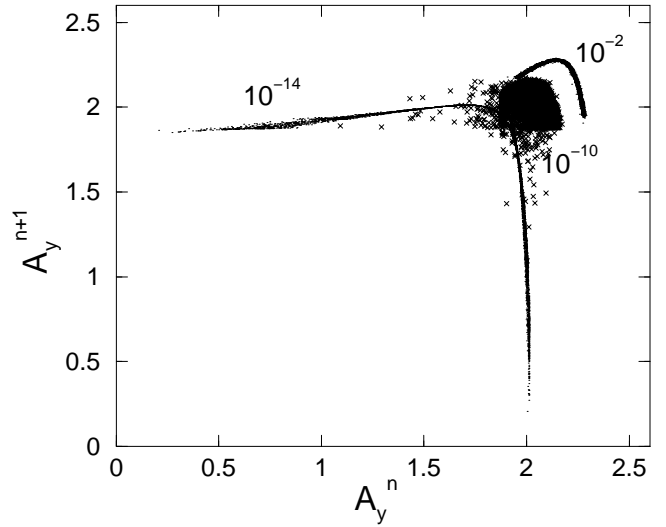}
\vspace{0.5cm}
\narrowtext
\caption{Return map constructed by successive maxima of $|y|$ for three different levels of noise
  $\sigma=10^{-14}$, $\sigma=10^{-10}$, and $\sigma=10^{-2}$.
}
\end{figure}

CR phenomenon in  the system is most appreciable  around the blowout bifurcation, i.e., in the on-off 
intermittency regime. In  general, the maximal  coherence is higher for the system above the bifurcation 
point.  If $c$ is far below the critical point, only large enough noise  can induce 
blowout motion from the subspace, and the coherence of the motion may have already  
destroyed by the  noise.
The peaks of $p_m$ and $\beta$ become lower and  narrower as $c$ decreases from $c=1.0$ and disappear 
when the subspace  becomes stable almost everywhere. 
For $c$ far above the critical point, the system has only seldom close  access  to the subspace, and 
weak noise has no significant effects on the motion, while strong noise 
 degrades the coherence of the motion away from the subspace. Typically,
one observes that both $p_m$ and $\beta$ keep unchanged for weak noise  and begin to decrease 
for high enough noise when the subspace becomes unstable almost everywhere.

\section{Conclusion}

In summary, we have shown that the phenomenon of coherence resonance  can be
naturally observed in some nonlinear dynamical systems possessing an invariant subspace, 
close to the  blowout bifurcation   
where previous studies  were often in very different context of on-off intermittency. A link
between these two distinct dynamical phenomena, CR and on-off intermittency, 
is the  fluctuation of the local transverse 
stability of the subspace due to the  oscillatory motion within the subspace. 
The noise-induced blowout motion manifests the coherent oscillation within the subspace, 
while 
the confinement of the nonlinearity reduces the chaotic fluctuation of the amplitude. 
The additive noise and the  
confinement of the nonlinearity to the transverse motion combine together to 
manifest and enhance  the intrinsic coherence
of the motion within the subspace to the maximal degree.  CR in this class of 
system with  mechanism  different from that in  previously studied  systems extends    
our understanding on nontrivial
and positive effects of noise on nonlinear dynamical systems, and could be  
physically and practically meaningful.

C. S. Zhou thanks Professor Degang Zhang for helpful discussion. 
This work is  supported in part by grants from the Hong Kong 
Research Grants Council (RGC) and the Hong Kong
Baptist University Faculty Research Grant (FRG).

\end{multicols}
\end{document}